# Strong tunability of epitaxial relationship and reconstruction at improper ferroelectric interface

Xin Li,[1†] Yu Yun,[1,4,†,*] Guodong Ren,[2] Arashdeep Singh Thind,[2] Amit Kumar Shah,[1] Rohan Mishra,[2,3] Xiaoshan Xu [1,5*]

[1] Department of Physics and Astronomy, University of Nebraska, Lincoln, Nebraska 68588, USA
[2] Institute of Materials Science & Engineering, Washington University in St. Louis, St. Louis MO, USA
[3] Department of Mechanical Engineering & Materials Science, Washington University in St. Louis, St. Louis MO, USA
[4] Department of Mechanical Engineering & Mechanics, Drexel University, Philadelphia, PA 19104-2875, USA
[5] Nebraska Center for Materials and Nanoscience, University of Nebraska, Lincoln, Nebraska 68588, USA

†These authors contributed equally

*Corresponding author: Y.Y. (yy549@drexel.edu)  and X.X. (xiaoshan.xu@unl.edu)

**Abstract**

The atomic structures at epitaxial film-substrate interfaces determine scalability of thin films and can result in new phenomena. However, it is challenging to control the interfacial structures since they are decided by the most stable atomic bonding. In this work, we report strong tunability of the epitaxial interface of improper ferroelectric hexagonal ferrites deposited on spinel ferrites. The selection of two interface types, related by a 90º rotation of in-plane epitaxial relations and featured by disordered and hybridized reconstructions respectively, can be achieved by growth conditions, stacking sequences, and spinel compositions. While the disordered type suppresses the primary $K_3$ structure distortion and ferroelectricity in hexagonal ferrites, the hybridized type is more coherent with the distortion with minimal suppression. This tunable interfacial structure provides critical insight on controlling interfacial clamping and may offer a solution for the long-standing problem of practical critical thickness in improper ferroelectrics.

Oxide interfaces continue to attract significant research interest as they enable emergent phenomena such as the juxtaposition of disparate properties, and the strong interfacial interactions resulting in new functionalities and device concepts[1,2]. The artificial interface between ferroelectric (or multiferroic) and ferromagnetic oxides provides an ideal arena for electric-field control of magnetism with reduced energy consumption such as a magnetoelectric spin–orbit (MESO) device [3,4]. Hexagonal ferrites (h-$R$FeO$_3$, $R$ = Sc and rare earth elements), a class of potential room-temperature multiferroic materials, offer opportunities to achieve atomic-scale engineering of ferroic orders in single-phase epitaxial films[5-10], superlattices[11-13], or single crystals[14], and have thus triggered widespread interest. Previous atomically-scale engineering efforts of h-$R$FeO$_3$ and the isomorphic hexagonal manganites (h-$R$MnO$_3$) have predominantly focused on the effects of epitaxial strain[7,8,15], chemical doping[8], or homo-type growth to modulate the multiferroic properties[16,17]. However, strong tunability of interfacial structure and ferroelectric critical thickness still remain critical challenges for the heteroepitaxial growth of h-$R$FeO$_3$, limiting potential interfacial magnetoelectric coupling for practical application[11,13].

Compared to conventional proper ferroelectrics wherein ferroelectricity is quenched below a critical thickness due to depolarization fields, the improper ferroelectric, h-$R$FeO$_3$, is expected to have no critical thickness since its ferroelectricity originates from non-polar K$_3$ mode (see Fig. 1(a)) [25,26], which can be represented by the order parameter $(Q, \phi)$, corresponding to the displacement amplitude and rotation angle of apical oxygen of FeO$_5$ bipyramids and the corrugation of the $R$ layers [27-29]. The polar $\Gamma_2^-$ mode which directly causes polarization is induced by the K3 mode (see Fig. 1(b)).

However, the corrugation of the first $R$ layer near the interface is usually suppressed (see Fig. 1 (c)) [7,17, 20], an effect called interfacial clamping. It leads to an experimentally observed critical thickness of h-$R$FeO$_3$ ultrathin films, even though theory predicts no critical thickness for free-standing films.[24] In addition to the ferroelectric critical thickness, to achieve strong magnetoelectric coupling at the h-$R$FeO$_3$/FM interface[13], persistence of the K$_3$ mode of h-$R$FeO$_3$ at the interface is necessary. In particular, the h-$R$FeO$_3$ (001)/spinel (111) heterostructure is promising for room-temperature interfacial multiferroicity and magnetoelectric effects due to the high Curie temperature for magnetic spinels such as CoFe$_2$O$_4$ and Fe$_3$O$_4$.[21,31-33] However, the structure of this heterointerface is poorly understood and its controllability has not been demonstrated [35].

In this work, two types of interfaces, corresponding to two different types of in-plane epitaxial relations, and two different types of reconstructions (type 1 disordered and type 2 hybridized) are revealed in both h-YbFeO$_3$(001)/CoFe$_2$O$_4$(111) heterostructure and h-LuFeO$_3$ (001)/CoFe$_2$O$_4$ (111) superlattices. The epitaxial relations of type 1 and type 2 h-$R$FeO$_3$/CoFe$_2$O$_4$ interfaces differ by a 90° in-plane rotation, which can be controlled by the growth temperature and stacking sequence. In the h-LuFeO$_3$ /CoFe$_2$O$_4$ superlattice, the consecutive interfaces alternate between type 1 and type 2, introducing a new degree of freedom for in-plane rotation within the superlattice. Based on a quantitative analysis of the spatial distribution of the order parameters, the type 2 interfacial reconstruction is found to effectively alleviate interfacial clamping, compared to the type 1 interface. These results reveal the general atomic-scale mechanism for maintaining bulk structural distortion at the heterointerface and open up the opportunity to resolve the long-standing detrimental interfacial clamping effect in ultrathin film of improper ferroelectrics.

Spinel family materials like CoFe$_2$O$_4$ (CFO) are promising candidates as substrate or buffer layer to grow single-phase epitaxial h-$R$FeO$_3$ films due to their crystal structure and growth conditions.[21, 28, 29] CFO is an inverse-spinel with a face-centered cubic (FCC) unit cell, in which Fe$^{3+}$ ions occupy tetrahedral (t) sites and both Fe$^{3+}$ and Co$^{2+}$ ions occupy octahedral (o) sites [36,37], corresponding to Fe$_t$(Co,Fe)$_o$O$_4$, as shown in Fig.1 (d). The primitive cell of the FCC structure can be chosen with a basis $\{\frac{\vec{a}-\vec{b}}{2}, \frac{\vec{b}-\vec{c}}{2}, \vec{b}\}$, where $\vec{a}$, $\vec{b}$, and $\vec{c}$ are the cubic coordinates; $\frac{\vec{a}-\vec{b}}{2}$ and $\frac{\vec{b}-\vec{c}}{2}$ form a triangular lattice in the (111) plane. The volume of the primitive cell is ¼ of that of the cubic unit cell, and the in-plane view of the primitive cell is shown in Fig. 1(e). Within this primitive cell, there is a layer of octahedral sites (octahedral layer) and a triple layer with tetrahedral/octahedral/tetrahedral stacking [31,38]. The schematic stackings of the octahedral layer and the triple layer, viewed along the [1-10] and the [11-2] directions, are displayed in Fig. 1(f).

The θ-2θ x-ray diffraction (XRD) of h-YbFeO$_3$ films grown on CFO (111) / LSMO (111) / STO (111) with different growth temperatures is shown in Fig. 2(a), where LSMO and STO stand for La$_{0.67}$Sr$_{0.33}$MnO$_3$ and SrTiO$_3$, respectively. Two types of in-plane epitaxial relationships can be identified by reflection high-energy electron diffraction (RHEED) images, as shown in Fig. 2(b), where the electron beam is along the LSMO [-211] direction. For the type 1 and type 2 in-plane relationships, h-YbFeO$_3$ [1-10] and h-YbFeO$_3$ [100] align with CFO [1-10], respectively. Compared with type 1, the relative in-plane orientation between h-YbFeO$_3$ and CFO is rotated by 90° in type 2, or 30° + n × 60° due to the six-fold rotational symmetry. Interestingly, type 1 can be stabilized when h-YbFeO$_3$ is first grown at a higher temperature (500°C) followed by annealing at 850 °C; the type 2 interface appears when h-YbFeO$_3$ is first grown at a lower temperature (300 °C) followed by annealing at 850 °C. This tunability suggests that the two types of in-plane epitaxial relationships of h-$R$FeO$_3$ (001)/ CFO (111) are close in energy, so a change of growth temperature by only 200 °C can cause such a dramatic change (90° rotation) of interfacial structure.

The θ-2θ XRD for the [h-LuFeO$_3$ (001) / CFO (111)]$_8$ superlattice grown on Al$_2$O$_3$ (001) substrate is shown in Fig. 3(a); the thickness of the h-LuFeO$_3$ layer and the CFO layer is 5.3 nm and 3 nm, respectively (see detailed growth condition in supplementary). Fig. 3(b) shows the x-ray reflectivity (XRR), in which the satellite peaks near the Bragg (00l) peaks indicate the formation of the superlattice; the inset shows the schematic layered structure of the superlattice. The in-plane epitaxial relationship is revealed by the RHEED images in Fig. 3(c), where the arrow indicates the stacking sequence of individual layers from bottom to top. The same two types of epitaxial relationships, as shown in Fig. 2(b), are also observed in Fig. 3 (c), but in an alternating fashion. The first h-LuFeO$_3$ (001) layer grown on CFO (111) adopts the type 2 relationship. Interestingly, the subsequent growth of CFO on h-LuFeO$_3$ adopts the type 1 relationship, which rotates 90° with respect to the previous CFO layer. The second h-LuFeO$_3$ layer on the rotated CFO adopts the type 2 relationship again; the next interface is the type 1. The schematics of the in-plane rotation of individual layers are given in Fig. 3(c), along with the RHEED images.

The alternating fashion suggests that the structure at the h-$R$FeO$_3$ (001) / CFO (111) interface is either determined by the corresponding CFO (111) / h-LuFeO$_3$ (001) interface underneath or by whether CFO is the overlayer or the underlayer. This observation confirms that the energetics of the h-$R$FeO$_3$ (001) / CFO (111) interface are subtle enough that, in addition to the deposition temperature, they can be controlled by the sequence of the deposited layers. Previously, we also studied epitaxial growth of h-$R$FeO$_3$ (001) films on the inverse-spinel, Fe$_3$O$_4$ (111), with $R$ = Lu and Yb. For h-$R$FeO$_3$ (001) deposited on Fe$_3$O$_4$ (111) at higher (750 °C) temperature followed by annealing, the type 2 in-plane epitaxial relationship was identified exclusively by RHEED studies[31]. This is in stark contrast to the h-$R$FeO$_3$ (001) / CFO (111) interface grown under similar conditions, for which the type 1 relationship dominates, as shown in Fig. 2(b). This comparison reveals that the composition of the spinel material is yet another variable for tuning the structure of the h-$R$FeO$_3$ (001) / spinel (111) interface.

To reveal the detailed interfacial structure at the h-$R$FeO$_3$ (001) / CFO (111) interface, we performed high-angle annular dark-field imaging using an aberration-corrected scanning transmission electron microscope (HAADF-STEM) for both type 1 and type 2 in-plane epitaxial relationships. Fig. 4(a) shows a HAADF-STEM image of the h-$R$FeO$_3$ (001) / CFO (111) interface with the type 1 in-plane relationship. The intensity of the atomic columns in a HAADF image is roughly proportional to the square of the average atomic number ($Z^2$)[42]. Thus, in h-YbFeO$_3$, the heavier Yb atomic columns appear brighter than the lighter Fe columns. On the h-YbFeO$_3$ side (above the interface), the corrugation of the Yb layer is obvious from the [100] viewing direction, and the magnitude of this corrugation determines the ferroelectric polarization. Moving closer to the interface, the corrugation gets suppressed. On the CFO side (below the interface), the octahedral layer / triple layer structure can be identified from the [-211] projection direction. The interface, indicated by the dotted box in the adjacent atomic model in Fig. 4(a), appears to consist of a Yb layer from h-YbFeO$_3$ on top, an octahedral layer from CFO at the bottom, and a middle layer of Fe/Co, highlighted by arrow, belongs to neither h-YbFeO$_3$ nor CFO. In particular, the middle layer features strong disorder that also propagates into the octahedral layer of CFO. This can be understood in terms of the significant differences in in-plane spatial periodicity. For example, in the horizontal direction of the image in Fig. 4(a), the spatial period is $\frac{\sqrt{3}}{2}a_H$=5.15 Å for h-YbFeO$_3$ and $\frac{\sqrt{2}}{2}a_S$=5.90 Å for CFO, where $a_H$=5.95 Å and $a_S$ = 8.35 Å are lattice constants of h-YbFeO$_3$ and CFO, respectively. Occasionally, interfaces with the same in-plane epitaxy relationship but without the middle layer are also observed (see Fig. S3). In that case, the disorder extends much deeper into the CFO side. Therefore, the type 1 interface features strong disorder due to the large mismatch between the periodicity of the h-YbFeO$_3$ and CFO in-plane lattices.

Fig. 4(b) shows a HAADF-STEM image of the h-$R$FeO$_3$ (001) / CFO (111) interface with the type 2 in-plane relationship. In this case, the interface (dotted box) consists of a Fe layer from h-YbFeO$_3$ on top, an octahedral layer from CFO at the bottom (see also Fig. S7), and a middle layer, highlighted by arrow, is composed of Yb and Fe/Co with a structure that belongs to neither h-YbFeO$_3$ nor CFO. In contrast to the strong disorder in type 1, type 2 is clearly ordered. In particular, the middle layer exhibits a Yb-Yb-Fe/Co pattern, corresponding to a reconstruction that features hybridization between CFO and h-YbFeO$_3$ structures. This ordered interface can be understood in terms of similar in-plane spatial periodicity. For example, in the horizontal direction of the image in Fig. 4(b), the spatial periodicity is $\frac{\sqrt{3}}{2}a_H$=5.15 Å for h-YbFeO$_3$ and $\frac{\sqrt{6}}{4}a_S$=5.11 Å for CFO. Previously, type 2 interface was observed in h-$R$FeO$_3$ (001) / Fe$_3$O$_4$ (111),[31] the triple layer of Fe$_3$O$_4$ (111) was proposed to neighbor the Fe layer of h-$R$FeO$_3$ due to the small in-plane mismatch discussed above. [31,33] However, in Fig. 4(b), the topmost layer of CFO part that can be identified as an octahedral layer. This may result from the high energy of the tetrahedral sites in the triple layer, which may undergo hybridization with the $R$ layer to form the Yb-Yb-Fe/Co pattern. The stability of the octahedral layer in CFO is in line with the widely studied perovskite-structure interface with a different-terminated monolayer [39-41].

Schematic in-plan views of the first rare-earth-containing layer from the h-$R$FeO$_3$ (001) / CFO (111) interface for the two types of in-plane relationships are displayed in Fig. 4(c) and (d). For type 1 interface, this layer is similar to those in bulk h-$R$FeO$_3$ except that the corrugation patterns are suppressed. On the other hand, for the type 2 interface, the hybridization of the $R$ layer and the triple layer in CFO leads to the periodic substitution of rare earth with Fe/Co. Indeed, the Fe/Co in this layer matches their positions in the tetrahedral site network in the triple layer of spinel [see Fig. 1(e)]. This substitution leads to a reduction of the HAADF-STEM intensity near interface, particularly when viewing from the h-$R$FeO$_3$ [1-10] (CFO [-211]) direction (see Fig. S5).

The two types of h-$R$FeO$_3$ (001) / spinel (111) interfaces can be summarized in terms of the layered structure of both h-$R$FeO$_3$ and spinel. Considering only the metal ions, the h-$R$FeO$_3$ structure is composed of alternating $R$ layers and Fe layers along the (001) direction, while the spinel structure consists of stackings of octahedral layers and triple layers along the (111) direction. For the type 1 in-plane epitaxial relationship, the Fe layer of h-$R$FeO$_3$ contacts the octahedral layer of CFO to form a (meta)stable structure. The largely mismatched spatial periodicity, however, causes strong disorder at the interface, especially in the Fe layer from h-$R$FeO$_3$. Energetically, type 1 is expected to have larger entropy due to the disorder, which is why it is favored at higher growth temperatures in h-$R$FeO$_3$ (001) / CFO (111). The type 2 in-plane epitaxial relationship corresponds to the (meta)stable structure in which the $R$ layer of h-$R$FeO$_3$ is in contact with the triple layer of CFO. In this case, the matching spatial periodicity minimizes the disorder. On the other hand, for the type 2 interface, the two layers hybridize to form a new layer with features from both h-$R$FeO$_3$ and CFO, where the $R$ and Fe/Co atoms retain their original positions in the bulk, as shown in Fig. 4(b). The revelation of the two types of h-$R$FeO$_3$ (001) / spinel (111) interfaces defies the common understanding of fixed in-plane epitaxial relationship at hetrointerface, and the strong tunability through control over the film-growth conditions and stacking sequence provides a way to alleviate the interfacial clamping, in addition to lattice matching scanerio.

As discussed above, either the Fe layer or the $R$ layer from h-$R$FeO$_3$ (001) can be selected to be in contact with the spinel substrate. A quantitative study of displacement of rare earth based on HAADF-STEM image can be used to extract the distribution of the order parameters ($Q$, $\phi$) [43,44]. Here, the $Q$ of K$_3$ structural distortion is related to the magnitude of corrugation of $R$ represented by $Q'$, where $Q$ is approximately proportional to $Q'$. For the type 1 interface, $Q'$ gradually increases with the increase of the $R$ layer number (see Fig. S6)[21]. Interestingly, for the type 2 interface, immediately after the initial reconstructed layer, $Q'$ reaches close to the bulk value of ≈ 30 pm, as shown in Fig. 5(a) and in the spatial mapping of $Q'$ in Fig. S7. In addition to the type 1 interface for h-YbFeO$_3$ on CFO discussed above, the type 1 interface for CFO on h-LuFeO$_3$ also induces interfacial clamping for h-LuFeO$_3$ in h-LuFeO$_3$/CFO superlattice, according to a previous STEM study[33]. Fig. 5(b) shows the quantitative comparison of $Q'$ in the initial rare earth layer of known hexagonal ferrite or manganite heterostructure interfaces[11,16,17,20,33,46-49]. For most interfaces involving hexagonal rare earth ferrites and manganites, due to the large mismatch between the in-plane lattice constants, the interfacial reconstruction features strong disorder, similar to the type 1 interface shown in Fig. 4(a). As a result, the initial rare earth layer suffers a strong reduction of $Q'$ (mostly <10 pm). For the interfaces where the in-plane lattice constants of hexagonal rare earth ferrites and manganites match closely with that of the substrate, the interfacial clamping effect can be much smaller. For example, in the h-LuFeO$_3$/LuFe$_2$O$_4$ superlattice, $Q'$ at the interface is close to the bulk value when h-LuFeO$_3$ is thick enough to exclude the electrostatic influence[11,44]. In addition, in the type 2 interface in Fig. 4(b), if the hybridized $R$/Fe/Co layer is not counted as part of the film (see related work of h-LuFeO$_3$ on SrCo$_2$Ru$_3$O$_{11}$(SCRO)[49]), there is virtually no clamping effect.

The effect of interfacial reconstruction on the $R$-layer corrugation near the interface can also be understood in terms of the displacement pattern of the K$_3$ distortion. As shown in Fig. 1(a), K$_3$ distortion in a bulk unit cell consists of the rotation of trigonal bipyramids and the corrugation of the $R$ layer with a propagation vector (1/3, 1/3, 0). If the neighboring trigonal bipyramid layer is disrupted, either by reducing the ordered rotation, or by (partly) removing the structure, the corrugation of the $R$ layer is expected to reduce. For h-$R$FeO$_3$ interface, if the initial termination layer is the rare earth, one neighboring trigonal bipyramid layer is absent. Combined with the lattice mismatch, the strong interfacial clamping effect (minimal <$Q'$>) is identified for h-YMnO$_3$/YSZ, h-LuFeO$_3$/YSZ, and h-YMnO$_3$/Al$_2$O$_3$. If the trigonal bipyramid layer is at the interface, <$Q'$> recovers but only partially, due to the reduction of ordered rotation, as exemplified in disordered type 1 h-YbFeO$_3$/CFO interface and h-LuFeO$_3$/Ir. On the other hand, in the type 2 interface of h-YbFeO$_3$/CFO, the first neighboring trigonal bipyramid layer and the ordered rotation are protected by the hybridized layer. Therefore, the interfacial clamping can be removed effectively, resulting in a much larger <$Q'$>. Ultimately, when the rare earth layer is on the trigonal bipyramid layer from $R$Fe$_2$O$_4$-like structure, such as h-LuFeO$_3$/Al$_2$O$_3$ and h-LuFeO$_3$/LuFe$_2$O$_4$, <$Q'$> is close to bulk value ($\approx$ 30 pm), or that of the interface from homo-type epitaxy[16,51].

In summary, two types of h-$R$FeO$_3$/spinel (111) interfaces with two in-plane epitaxy relationships connected by a 90° rotation have been observed. The rotation results in dramatically different lattice matching conditions between the in-plane lattice constants of the substrate and the film, leading to disordered and hybridized interfacial reconstructions. These, in turn, give rise to markedly different interfacial clamping effects on the initial rare earth layer. More importantly, the type of interface can be controlled by the initial growth temperature, stacking sequence, as well as spinel composition. These findings pave the way to achieve ultrathin improper ferroelectrics below 1 nm as well as potential room-temperature multiferroic composite.

**Acknowledgements**
This work is primarily supported by the National Science Foundation (NSF), grant DMR- 2419172. The work at Washington University was supported by the NSF through awards DMR-1806147 (AST) and DMR-2145797 (GR). The Microscopy work was conducted as part of a user project at the Center for Nanophase Materials Sciences (CNMS), which is a US Department of Energy, Office of Science User Facility at Oak Ridge National Laboratory. Research was partly performed in the Nebraska Nanoscale Facility: National Nanotechnology Coordinated Infrastructure and the Nebraska Center for Materials and Nanoscience (and/or NERCF), which are supported by the National Science Foundation under Award ECCS: 2025298, and the Nebraska Research Initiative.

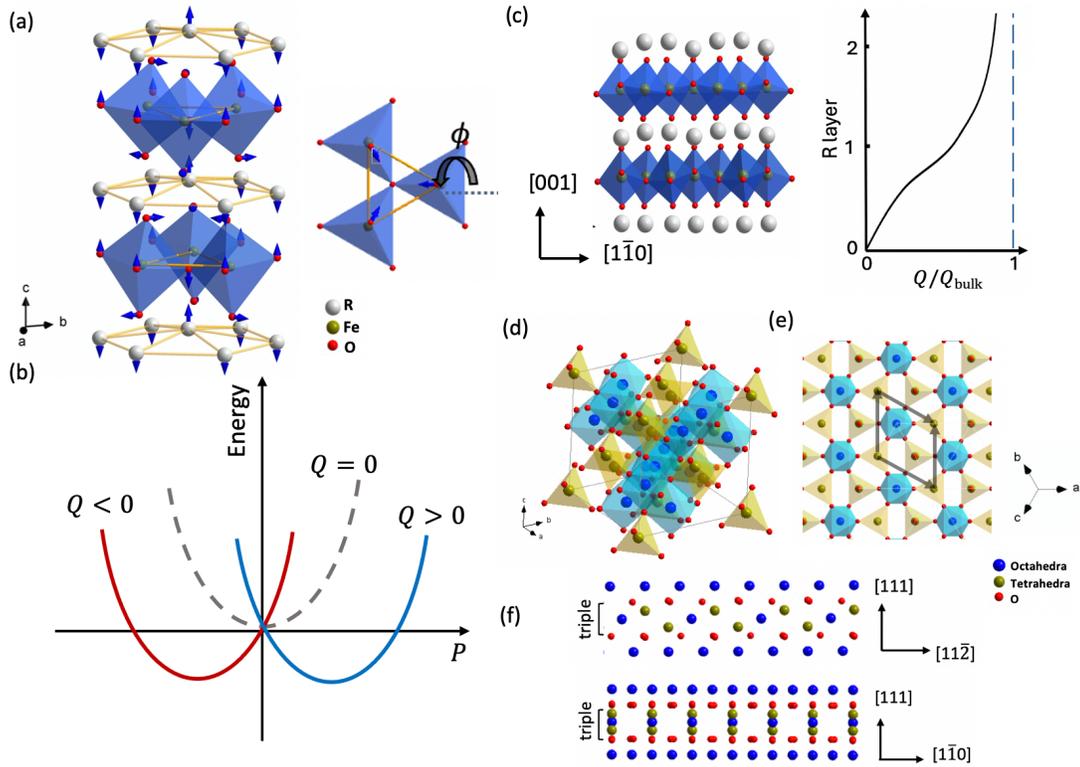

**Fig. 1** (a) The atomic structure of hexagonal ferrites, with arrows indicating the displacements of the $K_3$ structural distortion. (b) Schematic dependence of free energy on $\Gamma_2^-$ mode under the different $K_3$ distortion. (c) The schematic of the interfacial clamping effect, the right side is the atomic structure of hexagonal ferrites near the interface, viewed along [100]. (d) The atomic structure of inverse-spinel CFO, in which the yellow atoms are at the octahedral site, and the blue atoms are at the tetrahedral site. (e) The plane view of the primitive cell of CFO within the (111) plane. (f) Cross-section view of CFO atomic structure along $[1\bar{1}0]$ and $[11\bar{2}]$ directions.

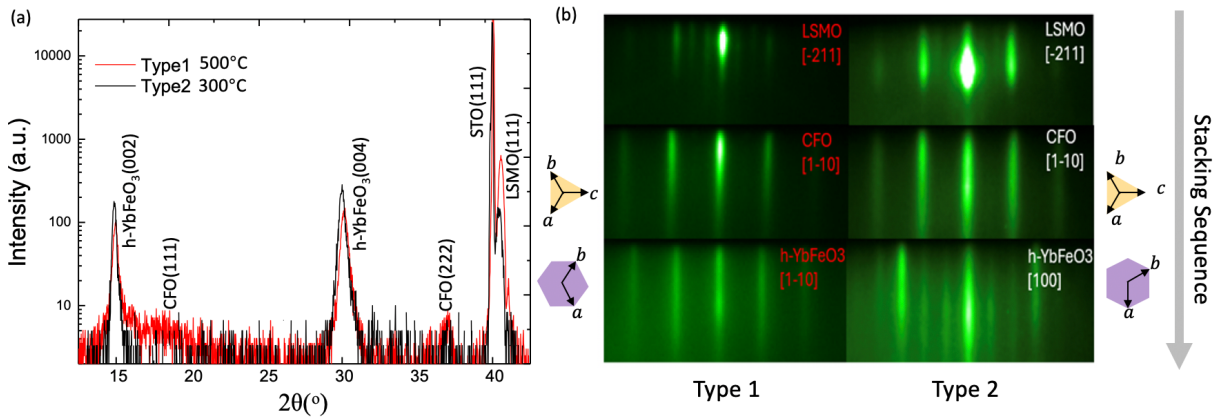

**Fig.2** (a) X-ray 2theta scan for 44 nm h-YbFeO$_3$ film with type 1 interface, and 45 nm h-YbFeO$_3$ film with type 2 interface on CFO/LSMO/STO (111), the temperature is for the growth temperature of the initial unit cell of h-YbFeO$_3$. (b) RHEED images for h-YbFeO$_3$ on CFO, with type 1 (left) and type 2 interfaces (right).

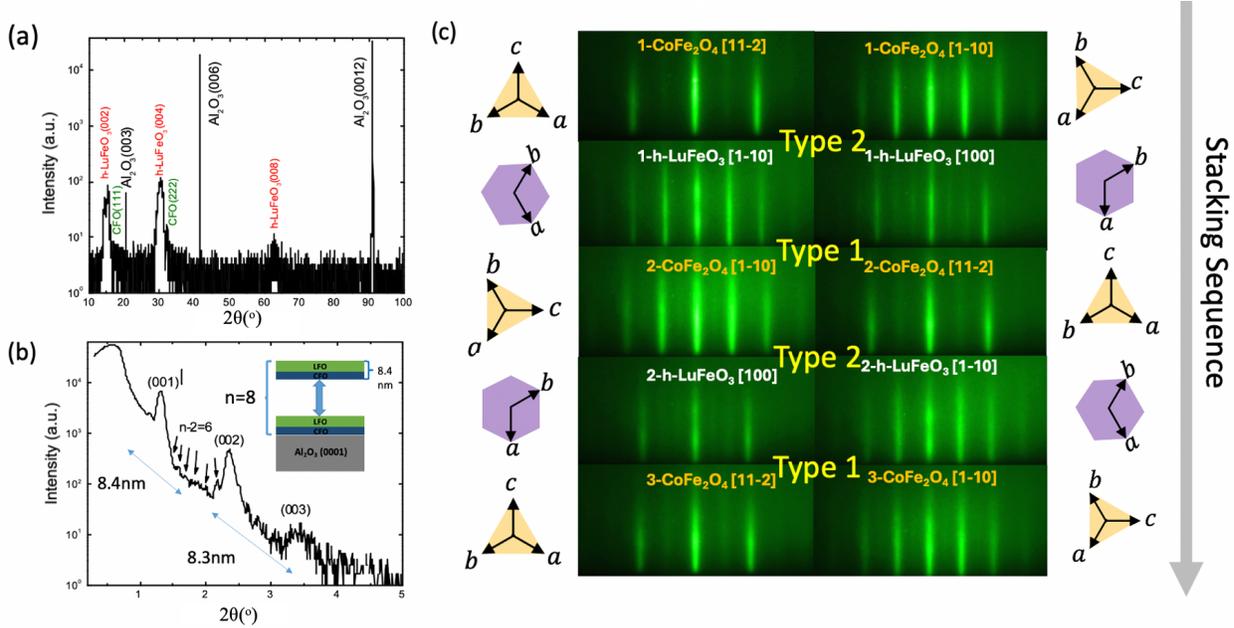

**Fig.3** (a) X-ray θ-2θ scan for (5.3nm h-LuFeO$_3$/3nm CFO)$_8$ superlattice and (b) related X-ray reflectivity. (c) The RHEED pattern for individual layer in superlattice. Left and right columns are the images taken with e-beams along two different directions respectively. The triangles and hexagons indicate the in-plane crystalline orientations.

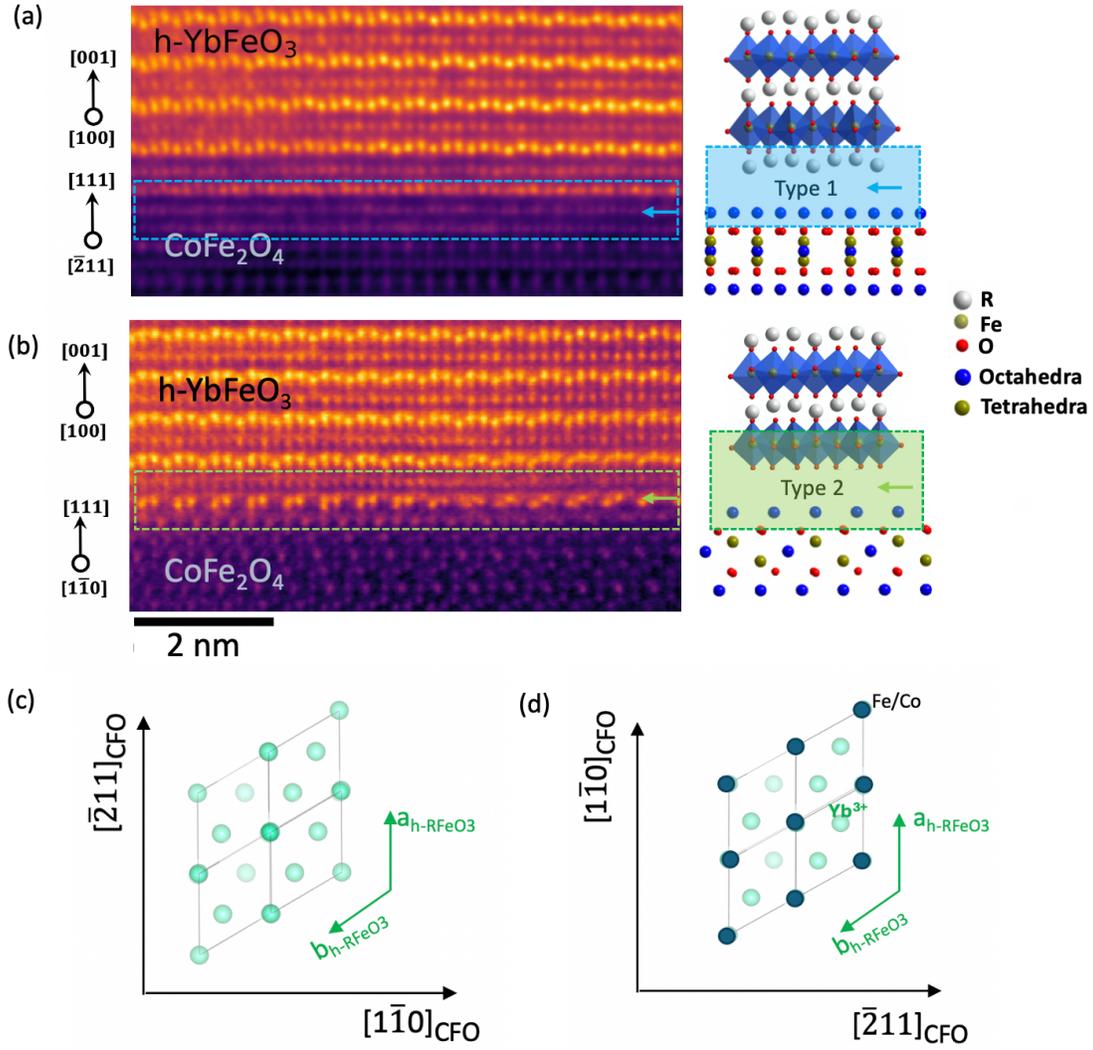

**Fig. 4** The HAADF-STEM image and schematic atomic structure for (a) type 1 and (b) type 2 interface in h-YbFeO$_3$/CFO/LSMO/STO (111). (c) and (d) are schematic in-plane atomic structures for the initial rare earth layer of h-*R*FeO$_3$ on CFO for type 1 and type 2 interface respectively.

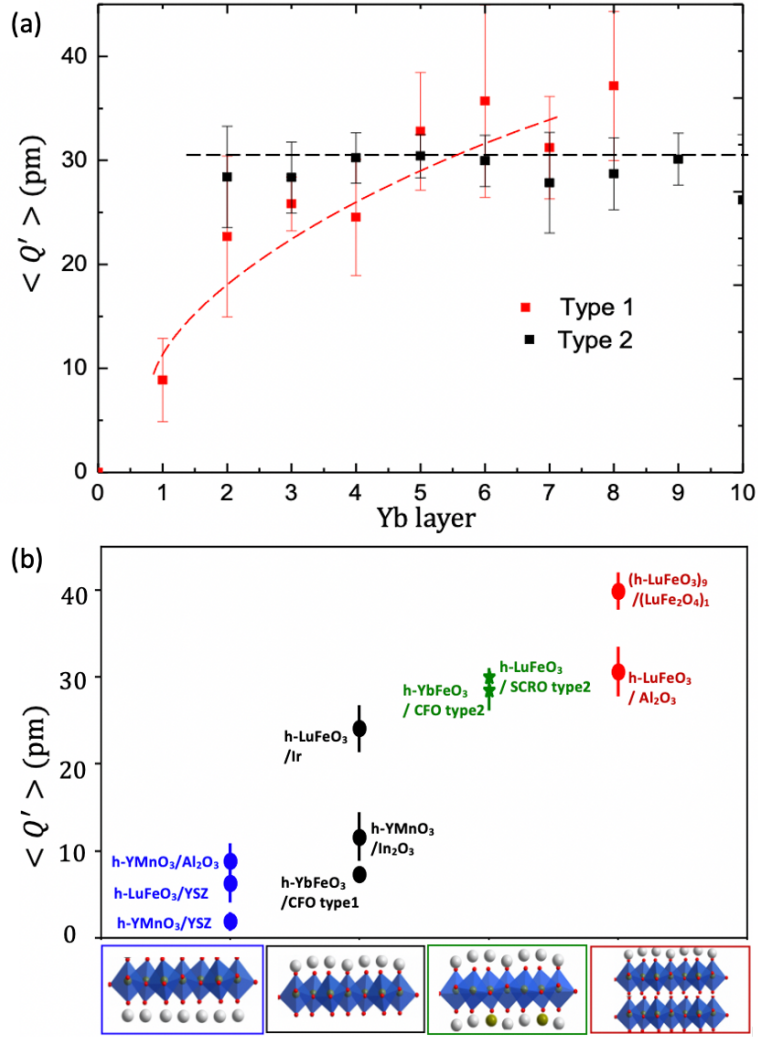

**Fig. 5** (a) The comparison of layer-dependent $Q'$ of h-YbFeO$_3$ on CFO with different type of interfacial reconstruction. (b) The summary of $Q'$ for initial rare-layers in h-$R$FeO$_3$ and h-$R$MnO$_3$ films with different interfaces[11,17, 20, 33, 46-49].